\documentstyle[amsmath,12pt]{article}

\begin{document}

\begin{titlepage}

\begin{center}

{\Large \bf
Exact Solutions of the Relativistic Many-Body Problem}
\vskip .6in

Domingo J. Louis-Martinez
\vskip .2in

Science One Program and\\
Department of Physics and Astronomy,\\ University of British
Columbia\\Vancouver, Canada

\end{center}

\vskip 3cm

\begin{abstract}
Exact solutions of the relativistic many-body problem are presented.
\end{abstract}
\end{titlepage}

In this paper we present exact solutions of the relativistic many-body
problem in action at a distance electrodynamics\cite{feynm}.

Let us consider the Fokker action:

\begin{equation}
S = \sum\limits^{}_{i} m_{i}c^{2} \int d\tau_{i} \eta_{\alpha\beta}
\dot{z}^{\alpha}_{i} \dot{z}^{\beta}_{i} +
\sum\limits^{}_{i} e_{i} \sum\limits^{}_{j\neq i} e_{j}
\int \int d\tau_{i} d\tau_{j}
\eta_{\alpha\beta} \dot{z}^{\alpha}_{i} \dot{z}^{\beta}_{j}
\delta\left((z_{i} - z_{j})^{2}\right)
\label{1}
\end{equation}

In (\ref{1}) , $m_{i}$ and $e_{i}$  $(i= 1,2,...,N)$ are the mass and
electric charge of particle $i$, $z^{\mu}_{i}$ its world line,
$\dot{z}^{\mu}_{i}$ its four-velocity, $\tau_{i}$ its proper time and
$c$ the speed of light. The metric tensor: $\eta_{\mu\nu}= 
diag(+1,-1,-1,-1)$.
The Dirac delta function in (\ref{1}) accounts for the interactions
propagating at the speed of light forward and backward in time\cite{feynm}.

The Minkowski equations of motion for N interacting relativistic electric
point-charges can be found from the action (\ref{1}) using the variational
principle\cite{barut}. We find:

\begin{equation}
m_{i}\ddot{z}^{\mu}_{i} = \sum\limits^{}_{s=-,+} K^{(s)\mu}_{i}
\label{A.1}
\end{equation}

\noindent where,

\begin{eqnarray}
K^{(s)\mu}_{i} & = & \frac{e_{i}}{2 c^{2}} \dot{z}_{i\nu}
\sum\limits^{}_{j \neq i}
\frac{e_{j}}
{|(z^{\sigma}_{i} - z^{(s)\sigma}_{j}) \dot{z}^{(s)}_{j\sigma}|^{3}}
\nonumber\\
& &  \left[\left((z^{\mu}_{i} - z^{(s)\mu}_{j})\ddot{z}^{(s)\nu}_{j}
- (z^{\nu}_{i} - z^{(s)\nu}_{j})\ddot{z}^{(s)\mu}_{j} \right)
\left(z^{\alpha}_{i} - 
z^{(s)\alpha}_{j}\right)\dot{z}^{(s)}_{j\alpha}\right.
\nonumber\\
& & - \left. \left((z^{\mu}_{i} - z^{(s)\mu}_{j})\dot{z}^{(s)\nu}_{j}
- (z^{\nu}_{i} - z^{(s)\nu}_{j})\dot{z}^{(s)\mu}_{j} \right)
\left(\left(z^{\alpha}_{i} - 
z^{(s)\alpha}_{j}\right)\ddot{z}^{(s)}_{j\alpha}
- 1\right)\right]
\label{A.2}
\end{eqnarray}

Here, $z^{(s)\mu}_{j} \equiv z^{\mu}_{j}(\tau^{(i,s)}_{j})$, $s=-,+$, are 
the
two roots of the equation:

\begin{equation}
\left(z_{i}(\tau_{i}) - z_{j}(\tau^{(i,s)}_{j})\right)^{2} = 0
\label{A.3}
\end{equation}

In this paper we show how the equations of motion (\ref{A.1},\ref{A.2}) 
can
be solved exactly, for any number of particles. In the solutions presented
here, the particles follow concentric uniform circular orbits:

\begin{equation}
z^{\mu}_{i} = \left(ct, r_{i}cos(\omega t + \phi_{i}),
r_{i}sin(\omega t + \phi_{i}), 0\right)
\label{2}
\end{equation}

Consider a particle $i$ at time $t$ in a reference frame $K$. Assume a 
signal
travelling at the speed of light is emitted from particle $j$ and reaches 
$i$
at time $t$.

$t - t^{(i,-)}_{j}$ is the time it takes for a signal to travel forward in
time at the speed of light from particle $j$ to particle $i$ in $K$.

$t^{(i,+)}_{j} - t$ is the time it takes for a signal to travel backward in
time at the speed of light from particle $j$ to particle $i$ in $K$.

The four-vector force acting on charge $i$ depends on the state of motion of
particle $i$ at time $t$ and, to account for the delay in the transmission
of the
interactions, on the states of motion of the remaining $N-1$ particles at
the past and future times $t^{(i,s)}_{j}  (j\neq i, s=-,+)$.

In action-at-a-distance electrodynamics, the interactions carry energy and
momentum to and from the particles and may simulate a field between them.
However, in the absence of electrically charged particles outside the 
system,
this fictitious field cannot carry energy or momentum into or away from the
system. If there are no other electrically charged particles in the 
universe,
the total energy and total momentum of a system of point particles are
conserved (assuming that only electromagnetic interactions take place).
In this paper we present exact formulas for the total mass and
total angular momentum (in the center-of-momentum frame) for $N$ interacting
relativistic point charges moving in concentric uniform circular orbits.

Consider the tangential and radial components of the net force acting on 
each
particle. Assuming the angular velocity $\omega$ to be constant, we find 
that
the tangential component of the net force acting on each charge $i$
$(i=1,2,...,N)$  vanishes:

\begin{eqnarray}
\sum\limits^{}_{j\neq i} e_{j} r_{j}\left[
\frac{\left(\left(1 - \frac{(r_{i}^{2} + r_{j}^{2}
- r_{i}r_{j}cos(\theta_{ij}))\omega^{2}}{c^{2}}\right)sin(\theta_{ij}) -
\frac{R^{ret}_{ij}\omega}{c}\left(cos(\theta_{ij})
- \frac{r_{i}r_{j}\omega^{2}}{c^{2}}\right)\right)}{\left(R^{ret}_{ij}
- \frac{r_{i}r_{j}\omega}{c}sin(\theta_{ij})\right)^{3}} + \right. & &
\nonumber\\
\left.
\frac{\left(\left(1 - \frac{(r_{i}^{2} + r_{j}^{2}
- r_{i}r_{j}cos(\eta_{ij}))\omega^{2}}{c^{2}}\right)sin(\eta_{ij}) +
\frac{R^{adv}_{ij}\omega}{c}\left(cos(\eta_{ij})
- \frac{r_{i}r_{j}\omega^{2}}{c^{2}}\right)\right)}{\left(R^{adv}_{ij}
+ \frac{r_{i}r_{j}\omega}{c}sin(\eta_{ij})\right)^{3}} \right] & = & 0
\label{3}
\end{eqnarray}

\noindent where,

\begin{equation}
\theta_{ij} = \phi_{i} - \phi_{j} + \frac{\omega R^{ret}_{ij}}{c}
\label{4.a}
\end{equation}

\begin{equation}
\eta_{ij} = \phi_{i} - \phi_{j} - \frac{\omega R^{adv}_{ij}}{c}
\label{4.b}
\end{equation}

\noindent and,

\begin{equation}
R^{ret}_{ij} = c \left(t - t^{(i,-)}_{j}\right)
\label{5.a}
\end{equation}

\begin{equation}
R^{adv}_{ij} = c \left(t^{(i,+)}_{j} - t\right)
\label{5.b}
\end{equation}

In other words, the tangential component of the net retarded force acting on
particle $i$ cancels with the tangential component of the net advanced force
acting on $i$.

The net work done by the electric force acting on each charge $i$
$(i=1,2,...,N)$ is zero.

Notice that from simple geometrical considerations, $R^{ret}_{ij}$ and
$R^{adv}_{ij}$ are determined by the equations:

\begin{equation}
\left(R^{ret}_{ij}\right)^{2} = r_{i}^{2} + r_{j}^{2} -
2 r_{i} r_{j} cos(\theta_{ij})
\label{6.a}
\end{equation}

\begin{equation}
\left(R^{adv}_{ij}\right)^{2} = r_{i}^{2} + r_{j}^{2} -
2 r_{i} r_{j} cos(\eta_{ij})
\label{6.b}
\end{equation}

For the radial component of the net force acting on charge $i$ we obtain:

\begin{eqnarray}
\frac{e_{i}}{2} \sum\limits^{}_{j\neq i} e_{j} \left[
\frac{(r_{i} - r_{j}cos(\theta_{ij}))
(1- \frac{r_{i}^{2}\omega^{2}}{c^{2}})(1- 
\frac{r_{j}^{2}\omega^{2}}{c^{2}})}
{\left(R^{ret}_{ij} - \frac{r_{i}r_{j}\omega}{c}sin(\theta_{ij})\right)^{3}}
+ \frac{\frac{\omega}{c} \left(- r_{j}sin(\theta_{ij}) +
\frac{r_{i}\omega}{c}R^{ret}_{ij}\right)}
{\left(R^{ret}_{ij} - \frac{r_{i}r_{j}\omega}{c}sin(\theta_{ij})\right)^{2}}
\right. & &\nonumber\\
\left. + \frac{(r_{i} - r_{j}cos(\eta_{ij}))
(1- \frac{r_{i}^{2}\omega^{2}}{c^{2}})(1- 
\frac{r_{j}^{2}\omega^{2}}{c^{2}})}
{\left(R^{adv}_{ij} + \frac{r_{i}r_{j}\omega}{c}sin(\eta_{ij})\right)^{3}}
+ \frac{\frac{\omega}{c} \left(r_{j}sin(\eta_{ij}) +
\frac{r_{i}\omega}{c}R^{adv}_{ij}\right)}
{\left(R^{adv}_{ij} + \frac{r_{i}r_{j}\omega}{c}sin(\eta_{ij})\right)^{2}}
\right] & & \nonumber\\
= - \frac{m_{i}r_{i}\omega^{2}}
{\left(1 - \frac{r_{i}^{2}\omega^{2}}{c^{2}}\right)^{\frac{1}{2}}}
\hspace{10cm} & &
\label{7}
\end{eqnarray}

The total mass of the system can be found following the method of Landau and
Lifshitz \cite{landau}. We find:

\begin{equation}
M = \sum\limits^{N}_{i=1} m_{i}
\left(1 - \frac{r_{i}^{2} \omega^{2}}{c^{2}}\right)^{\frac{1}{2}}
\label{8}
\end{equation}

As a conjecture, I present here a formula for the total angular momentum of
the system (in the center-of-momentum inertial reference frame):

\begin{equation}
L = - \frac{1}{4\omega} \sum\limits^{}_{i} \sum\limits^{}_{j \neq i}
e_{i} e_{j} \left[
\frac{\left(1 - \frac{r_{i}r_{j}\omega^{2}}{c^{2}}cos(\theta_{ij})\right)}
{\left(R^{ret}_{ij} - \frac{r_{i}r_{j}\omega}{c}sin(\theta_{ij})\right)} +
\frac{\left(1 - \frac{r_{i} r_{j} \omega^{2}}{c^{2}}cos(\eta_{ij})\right)}
{\left(R^{adv}_{ij} + \frac{r_{i}r_{j}\omega}{c}sin(\eta_{ij})\right)}
\right]
\label{9}
\end{equation}

Formula (\ref{9}) gives the correct expression for the total angular 
momentum
of N interacting point-charges in uniform circular motion if the velocities
of all the particles are small compared with the velocity of light (I have
checked this up to terms of second order, $\frac{v^{2}}{c^{2}}$).

For $N=2$, Eqs(\ref{3},\ref{7},\ref{8},\ref{9}) reduce
to the exact solutions found by Schild \cite{schild} for the electromagnetic
two-body problem.

From (\ref{4.a},\ref{4.b},\ref{6.a},\ref{6.b}) it follows that:

\begin{equation}
R^{adv}_{ji} = R^{ret}_{ij}
\label{10}
\end{equation}

\begin{equation}
\eta_{ji} = - \theta_{ij}
\label{11}
\end{equation}

Therefore, formula (\ref{9}) can be written in a more compact form.

It is also not difficult to prove that Eqs (\ref{3}) are not all 
independent.
Indeed, from (\ref{10},\ref{11}) it follows that:

\begin{eqnarray}
\sum\limits^{}_{i}\sum\limits^{}_{j\neq i} (e_{i} r_{i}) (e_{j} r_{j})
\left[ \frac{\left(\left(1 - \frac{(r_{i}^{2} + r_{j}^{2}
- r_{i}r_{j}cos(\theta_{ij}))\omega^{2}}{c^{2}}\right)sin(\theta_{ij}) -
\frac{R^{ret}_{ij}\omega}{c}\left(cos(\theta_{ij})
- \frac{r_{i}r_{j}\omega^{2}}{c^{2}}\right)\right)}{\left(R^{ret}_{ij}
- \frac{r_{i}r_{j}\omega}{c}sin(\theta_{ij})\right)^{3}} + \right.
\nonumber\\
\left.
\frac{\left(\left(1 - \frac{(r_{i}^{2} + r_{j}^{2}
- r_{i}r_{j}cos(\eta_{ij}))\omega^{2}}{c^{2}}\right)sin(\eta_{ij}) +
\frac{R^{adv}_{ij}\omega}{c}\left(cos(\eta_{ij})
- \frac{r_{i}r_{j}\omega^{2}}{c^{2}}\right)\right)}{\left(R^{adv}_{ij}
+ \frac{r_{i}r_{j}\omega}{c}sin(\eta_{ij})\right)^{3}} \right]
\label{12}
\end{eqnarray}

\noindent identically vanishes (since the expression inside the bracket is
antisymmetric).

Eqs (\ref{3},\ref{7}) completely determine the $2N - 1$ unknows $r_{i}$
$(i=1,2,...,N)$, $\phi_{i} - \phi_{i-1}$ $(i=2,...,N)$, as functions of
the angular velocity $\omega$.

Let us now consider two simple classes of solutions for N interacting point
particles moving along circular orbits:

{\it Regular polygonal solutions with a nucleus:}

Assume all the masses are equal $(m_{i} = m)$ and all charges are equal
$(e_{i} = - e)$ $(i = 1,2,...,N)$. Assume there is a central mass $m_{0}$ 
with
electric charge $e_{0} = N e$.

The charges $e_{i}$ $(i=1,2,...,N)$ are located at the vertices of a regular
polygon:

\begin{equation}
\phi_{j+1} - \phi_{j} = \frac{2\pi}{N}
\label{13}
\end{equation}

\begin{equation}
\frac{m\omega^{2}r^{3}}
{e^{2}\left(1 - \frac{r^{2}\omega^{2}}{c^{2}}\right)^{\frac{1}{2}}} =
N - \sum\limits^{N-1}_{k=1}
\frac{\left((1+\frac{r^{4}\omega^{4}}{c^{4}})(1-cos\gamma_{k}) +
\frac{r^{2}\omega^{2}}{c^{2}}sin^{2}\gamma_{k}
- \frac{l_{k}\omega}{c}\left(1+\frac{r^{2}\omega^{2}}{c^{2}}\right)
sin\gamma_{k}\right)}
{\left(\frac{l_{k}}{r} - \frac{r\omega}{c}sin\gamma_{k}\right)^{3}}
\label{14}
\end{equation}

\noindent where,

\begin{equation}
\gamma_{k} = \frac{2\pi}{N} k + \frac{\omega l_{k}}{c}
\label{15}
\end{equation}

\begin{equation}
l_{k} = \sqrt{2} r \left(1 - cos(\gamma_{k})\right)^{\frac{1}{2}}
\label{16}
\end{equation}

$(k = 1,2,...,N-1)$

For the total mass and total angular momentum we find:

\begin{equation}
M = m_{0} + N m \left(1 - \frac{r^{2}\omega^{2}}{c^{2}}\right)^{\frac{1}{2}}
\label{17}
\end{equation}

\begin{equation}
L = \frac{N^{2}e^{2}}{\omega r} \left[
1 - \frac{1}{2 N}  \sum\limits^{N-1}_{k=1}
\frac{\left(1 - \frac{r^{2}\omega^{2}}{c^{2}}cos\gamma_{k}\right)}
{\left(\frac{l_{k}}{r} - \frac{r\omega}{c}sin\gamma_{k}\right)}
\right]
\label{18}
\end{equation}

{\it Regular polygonal solutions without a nucleus:}

Assume all the masses are equal, $m_{i} = m$, and the charges all have the
same magnitude. Half of the particles are positively charged and the other
half negatively charged: $e_{i} = (-1)^{i} e$, $i=1,2,...,N$. Here we assume
the number of particles $N$ is even. The charges are located at the vertices
of a regular polygon, with alternating signs (nearest neighbours attract
each other, as their charges have opposite signs). We find for this class
of solutions:

\begin{equation}
\phi_{j+1} - \phi_{j} = \frac{2\pi}{N}
\label{19}
\end{equation}

\begin{equation}
\frac{m\omega^{2}r^{3}}
{e^{2}\left(1 - \frac{r^{2}\omega^{2}}{c^{2}}\right)^{\frac{1}{2}}} =
\sum\limits^{N-1}_{k=1} (-1)^{k+1}
\frac{\left((1+\frac{r^{4}\omega^{4}}{c^{4}})(1-cos\gamma_{k}) +
\frac{r^{2}\omega^{2}}{c^{2}}sin^{2}\gamma_{k}
- \frac{l_{k}\omega}{c}\left(1+\frac{r^{2}\omega^{2}}{c^{2}}\right)
sin\gamma_{k}\right)}
{\left(\frac{l_{k}}{r} - \frac{r\omega}{c}sin\gamma_{k}\right)^{3}}
\label{20}
\end{equation}

\noindent where $\gamma_{k}$ and $l_{k}$ $(k=1,...,N-1)$ are determined by
the relations (\ref{15}, \ref{16}).

For the total mass and total angular momentum we find:

\begin{equation}
M = N m \left(1 - \frac{r^{2}\omega^{2}}{c^{2}}\right)^{\frac{1}{2}}
\label{21}
\end{equation}

\begin{equation}
L = \frac{Ne^{2}}{2\omega r} \sum\limits^{N-1}_{k=1} (-1)^{k+1}
\frac{\left(1 - \frac{r^{2}\omega^{2}}{c^{2}}cos\gamma_{k}\right)}
{\left(\frac{l_{k}}{r} - \frac{r\omega}{c}sin\gamma_{k}\right)}
\label{22}
\end{equation}

\end{document}